# Magnetic Field Generation in Planets and Satellites by Natural Nuclear Fission Reactors

by

J. Marvin Herndon
Transdyne Corporation
San Diego, CA 92131 USA

27 July 2007

**Abstract:** One of the most fundamental problems in physics has been to understand the nature of the mechanism that generates the geomagnetic field and the magnetic fields of other planets and satellites. For decades, the dynamo mechanism, thought to be responsible for generating the geomagnetic field and other planetary magnetic fields, has been ascribed to convection in each planet's iron-alloy core. Recently, I described the problems inherent in Earth-core convection and proposed instead that the geomagnetic field is produced by convection in the electrically conductive, fluid, fission-product sub-shell of a natural nuclear fission reactor at the center of the Earth, called the georeactor. Here I set forth in detail the commonality in the Solar System of the matter like that of the inside of the Earth, which is my basis for generalizing the concept of planetary magnetic field generation by natural planetocentric nuclear fission reactors.

## Introduction

Currently active internally generated magnetic fields have been detected in six planets (Mercury, Earth, Jupiter, Saturn, Uranus, and Neptune) and in one satellite (Jupiter's moon Ganymede). Magnetized surface areas of Mars and the Moon indicate the former existence of internally generated magnetic fields in those bodies. The purpose of this communication is to suggest that those magnetic fields arise from the same georeactor-type mechanism which I have suggested generates and powers the Earth's magnetic field [1].

There is clear evidence that certain planets contain internal energy sources. In 1969 astronomers discovered that Jupiter radiates into space more energy than it receives from the Sun. Verification followed, indicating that not only Jupiter, but Saturn and Neptune as well each radiate approximately twice as much energy as they receive from the Sun [2, 3]. For two decades planetary scientists thought that they had considered and eliminated known planetary-scale energy sources, declaring "by default" or "by elimination" the observed internal must be a relic, leftover energy from planetary formation about 4.5 billion years ago [4, 5].

Applying Fermi's nuclear reactor theory [6], in 1992 I demonstrated the feasibility for planetocentric nuclear fission reactors as the internal energy sources for the gas-giant outer planets [7]. Initially, I considered only hydrogen-moderated thermal neutron reactors, but shortly demonstrated the feasibility for fast neutron breeder reactors as well, which admitted the possibility of planetocentric nuclear reactors in planets like Earth [8-10].

It is known that the Earth has an internal energy source at or near its center that powers the mechanism which generates and sustains the geomagnetic field. Applying Fermi's nuclear reactor theory [6], I demonstrated the feasibility of a planetocentric nuclear fission reactor as the energy source for the geomagnetic field [8]. Subsequent state-of-the-art numerical simulations, made at Oak Ridge National Laboratory, verified my conjecture that the georeactor could indeed function over the lifetime of the Earth as a fast neutron breeder reactor and, significantly, would produce helium in the same range of isotopic compositions observed in oceanic basalts [11-13]. At this point, though, I had only considered planetocentric nuclear fission reactors as planetary energy sources, not as mechanisms for generating planetary magnetic fields. Recently, though, I suggested that the georeactor is responsible, not only for powering the geomagnetic field, but for also being the mechanism responsible for actually generating the geomagnetic field [1].

In this paper is I suggest that the mechanism for generating planetary and satellite magnetic fields and for providing their requisite energy are one and the same, planetocentric nuclear fission reactors, like the Earth's georeactor [1]. That generalization is based upon fundamental considerations demonstrating the commonality of highly-reduced, deep-Earth type matter, particularly within massive-cored planets of our Solar System.

**Nature of Planetary Matter**

Only three processes, operant during the formation of the Solar System, are responsible for the diversity of matter in the Solar System and are directly responsible for planetary internal-structures, including planetocentric nuclear fission reactors, and for dynamical processes, including and especially, geodynamics. These processes are: (*i*) Low-pressure, low-temperature condensation from solar matter in the remote reaches of the Solar System or in the interstellar medium, which leads to oxygen-rich condensate; (*ii*) High-pressure, high-temperature condensation from solar matter associated with planetary-formation by raining out from the interiors of giant-gaseous protoplanets, which leads to oxygen-starved planetary interiors, and; (*iii*) Stripping of the primordial volatile components from the inner portion of the Solar System by super-intense solar wind associated with T-Tauri phase mass-ejections, presumably during the thermonuclear ignition of the Sun [14].

The constancy in isotopic compositions of most of the elements of the Earth, the Moon, and the meteorites indicates formation from primordial matter of common origin. Primordial elemental composition is yet evident to a great extent in the photosphere of the Sun and, for the less volatile, rock-forming elements, in chondrite meteorites. There is, however, a fundamental degree of complexity which has posed an impediment to understanding for more than half a century: Instead of just one type of chondrite there are three, with each type characterized by its own strikingly unique state of oxidation. Understanding the nature of the processes that yielded those three distinct types of matter from one common progenitor forms the basis for understanding much about planetary formation, their compositions, and the processes they manifest, including magnetic field production

Five major elements [Fe, Mg, Si, O, and S] comprise at least 95% of the mass of each chondrite and, by implication, each of the terrestrial planets, and act as a buffer assemblage. Minor and trace elements are slaves to that buffer system and are insufficiently abundant to alter oxidation state. For decades, the abundances of major rock-forming elements ($E_i$) in chondrites have been expressed in the literature as ratios, usually relative to silicon ($E_i$/Si) and occasionally relative to magnesium ($E_i$/Mg). By expressing Fe-Mg-Si elemental abundances as molar (atom) ratios relative to iron ($E_i$/Fe), as shown in Figure 1, I discovered a fundamental relationship bearing on the nature of chondrite matter which has fundamental implications on the nature of planetary matter in our Solar System [15].

In Figure 1, chondrite data points scatter about three distinct, well defined, least squares fit, straight lines, unique to their classes, despite mineralogical differences observed among members within a given class of chondrites.

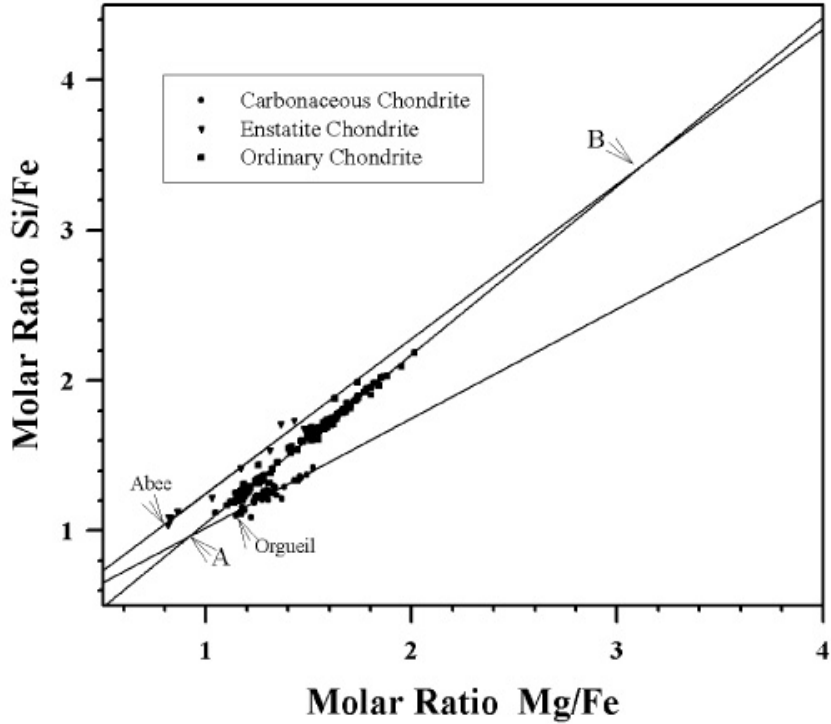

**Figure 1.** Whole-rock major element ratios for 10 enstatite chondrites, 39 carbonaceous chondrites, and 157 ordinary chondrites plot as three well defined straight lines. For details on statistics and implications see [15].

At one level of understanding, Figure 1 means that the well-mixed primordial matter became, or evolved to become, only three distinct types of matter which still retain more-or-less the full complement of readily condensable elements and which became planetary building materials. At a deeper level, though, the relationship shown in Figure 1 admits the possibility of ordinary chondrites having been derived from mixtures of two components, representative of the other two types of matter, mixtures of a relatively undifferentiated carbonaceous-chondrite-like *primitive* component and a partially differentiated enstatite-chondrite-like *planetary* component, which I have suggested might be comprised of matter stripped from the protoplanet of Mercury, presumably by the T-Tauri solar wind during thermonuclear ignition of the Sun [15]. In other words, ordinary chondrite matter is not a primary building material for planets, although it might contribute a veneer to the terrestrial planets, especially to Mars.

Much confusion has arisen from decades of making computational models which erroneously assume that the mineral assemblage characteristic of ordinary chondrites formed in equilibrium in an atmosphere of solar composition at very low pressures, ca. $10^{-4}$ to $10^{-5}$ bars, and that ordinary-chondrite-like matter comprises planetary interiors.

I have shown that ordinary chondrite formation necessitates, not an atmosphere of solar composition, but instead an atmosphere depleted in hydrogen by a factor of about 1000 [16] and depleted somewhat in oxygen relative to solar matter [17]. Moreover, from Figure 1, the ordinary chondrites appear, not primary, but rather as a secondary mixture, leaving only two types of primary matter, the oxygen-rich carbonaceous chondrite-type matter and the oxygen-starved enstatite chondrite-type matter.

As early as 1940, scientists, including the renowned Harvard geophysicist Francis Birch, built geophysics upon the premise that the Earth is like an ordinary chondrite, one of the most common types of meteorites observed impacting Earth, while totally ignoring another, albeit less abundant type, called enstatite chondrites. As I discovered in 1980, if the Earth is indeed like a chondrite meteorite as widely believed for good reasons, Earth is like an enstatite chondrite, not an ordinary chondrite [18]. Imagine melting a chondrite in a gravitational field. At elevated temperatures, the iron metal and iron sulfide components will alloy together, forming a dense liquid that will settle beneath the silicates like steel on a steel-hearth. The Earth is like a spherical steel-hearth with a fluid iron-alloy core surrounded by a silicate mantle.

The Earth's core comprises about 32.5% of the planets mass. Only the enstatite chondrites, not the ordinary chondrites, have the sufficiently high proportion of iron-alloy that is observed for the core of the Earth, as shown in Figure 2. Moreover, as I discovered, components of the interior of the Earth can be identified with corresponding components of an enstatite chondrite meteorite: (1) The inner core being nickel silicide; (2) Earth-core precipitates CaS and MgS at the core-mantle boundary; (3) The lower mantle consisting of essentially FeO-free $MgSiO_3$; and, (4) The boundary between the upper and lower mantle being a compositional boundary with the matter below that boundary, the endo-Earth, being like an enstatite chondrite [18-21]. Those discoveries and insights led to a fundamentally different view of Earth formation, dynamics, energy production, and energy transport process [14, 22, 23].

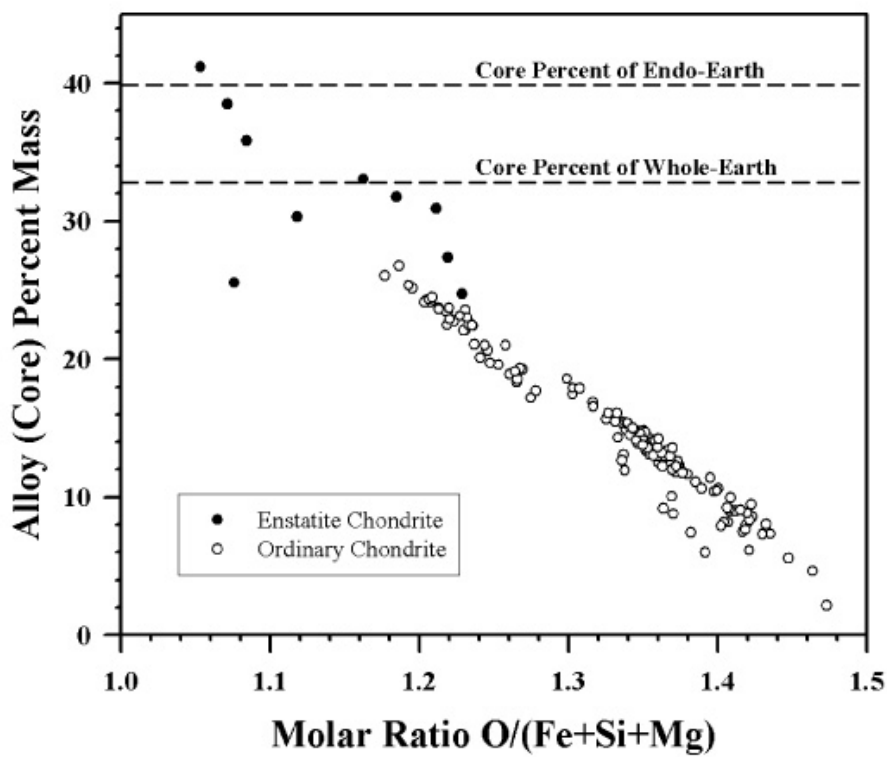

**Figure 2.** The percent alloy (mainly iron metal plus iron sulfide) of 157 ordinary chondrites and 9 enstatite chondrites plotted against a measure of oxygen content. The Earth as a whole, and especially the endo-Earth (core plus lower mantle) is like an enstatite chondrite and unlike an ordinary chondrite. For additional information, see [14, 21].

In the 1940s and 1950s, the idea was generally discussed about planets "raining out" from inside of giant gaseous protoplanets with hydrogen gas pressures on the order of $10^2$-$10^3$ bars [24-27]. But, in the early 1960s, scientists instead began thinking of primordial matter, not forming dense protoplanets, but rather spread out into a very low-density "solar nebula" with hydrogen gas pressures on the order of $10^{-4}$ to $10^{-5}$ bars. The idea of low-density planetary formation, often referred to as the "standard model", envisioned that dust would condense at fairly low temperatures, and then would gather into progressively larger grains, and become rocks, then planetesimals, and ultimately planets [28, 29].

These two ideas about planetary formation embody fundamentally different condensation processes which are the underlying cause for the two unique primary types of chondritic matter shown in Figure 1. The immediate implication is that both processes were operant during the formation of the Solar System. The relative extent and region of each process can be ascertained to some certitude from thermodynamic considerations together with planetary data. Even within present limitations, a consistent picture emerges that is quite unlike the so-called "standard model of solar system formation" [14].

From thermodynamic considerations it is possible to make some generalizations related to the condensation process in an atmosphere of solar composition. At the foundation, there are two dominant considerations, one essentially independent of pressure and one a strong function of pressure which are responsible for formation of the two primary types of Solar System matter.

In an atmosphere of solar composition, oxygen fugacity is dominated by the gas-phase reaction $H_2 + \frac{1}{2}O_2 = H_2O$ which is a function of temperature, but is essentially independent of pressure over a wide range of pressures where ideal gas behavior is approached. Oxygen fugacity controls the condensate state of oxidation at a particular temperature. At high temperatures the state of oxidation is extremely reducing, while at low temperatures it is quite oxidizing. The state of oxidation of the condensate ultimately becomes fixed at the temperature at which reaction with the gas phase ceases and/or equilibrium is frozen-in by the separation of gases from the condensate.

Condensation of an element or compound is expected to occur when its partial pressure in the gas becomes greater than its vapor pressure. Generally, at high pressures in solar matter, condensation is expected to commence at high temperatures, while at low pressures, such as $10^{-4}$ to $10^{-5}$ bar, condensation is expected to progress at relatively low temperatures at a fairly oxidizing range of oxygen fugacity. At low temperatures, all of the major elements in the condensate may be expected to be oxidized because of the great abundance of oxygen in solar matter relative to the other major condensable elements [30]. Beyond these generalizations, in this low-pressure regime, precise theoretical predictions of specific condensate compounds may be limited by kinetic nucleation dynamics and by gas-grain temperature differences arising because of the different mechanisms by which gases and condensate lose heat.

Among the thousands of known chondrites, only a few, like the famous Orgueil carbonaceous chondrite, have a state of oxidation and mineral components with characteristics similar to those expected as a condensate from solar matter at low pressures. Essentially all of the major elements in these few chondrites are oxidized, including sulfur.

The idea of planetary formation from a diffuse solar nebula, with hydrogen pressures on the order of $10^{-4}$ to $10^{-5}$ bar, envisioned that dust would condense at fairly low temperatures, and then would gather into progressively larger grains, and become rocks, then planetesimals, and ultimately planets. In the main, that picturesque idea leads to the contradiction of the terrestrial planets having insufficiently massive cores, because the condensate would be far too oxidized for a high proportion of iron metal to exist. But as evidenced by Orgueil and similar meteorites, such low-temperature, low-pressure condensation did in fact occur, perhaps only in the evolution of matter of the outer regions of the Solar System or in interstellar space, and thus may contribute to terrestrial planet formation only as a component of late-addition planetary veneer.

On the basis of thermodynamic considerations, Eucken suggested in 1944 core-formation in the Earth as a volatility-based consequence of successive condensation from solar matter from the central region of a hot, gaseous protoplanet with molten iron metal first raining out at the center [24]. Except for a few investigations initiated in the 1950s early 1960s [25, 26, 31, 32], that idea languished when interest was diverted to Cameron's low-pressure solar nebula models [33].

On the basis of thermodynamic considerations, Suess and I showed at the high-temperatures for condensation at high-pressures, solar matter is sufficiently reducing, *i.e.*, it has a sufficiently low oxygen fugacity, for the stability of some enstatite chondrite minerals [34]. However, formation of enstatite-chondrite-like condensate would necessitate thermodynamic equilibrium being frozen-in at near-formation temperatures. At present, there is no adequate published theoretical treatment of solar-matter condensation from near the triple-point. But from thermodynamic and metallurgical considerations, some generalizations can be made. At the high temperatures at which condensation is possible at high pressures, nearly everything reacts with everything else and nearly everything dissolves in everything else. At such pressures, molten iron, together with the elements that dissolve in it, is the most refractory condensate.

From solar abundances [30], the calculated mass of protoplanetary-Earth was 275-305$m_E$, not very different from the mass of Jupiter, 318$m_E$. The formation of early-phase close-in gas giants in our own planetary system is consistent with observations and implications of near-to-star giant gaseous planets in other planetary systems [35-37]. It is thus reasonable to expect that the giant planets possess interior rock-plus-alloy kernels of enstatite-chondritic-like matter.

In the absence of evidence to the contrary, the observed enstatite-chondritic composition of the terrestrial planets, as indicated by their massive cores, permits the deduction that these planets formed by raining out from the central regions of hot, gaseous protoplanets [14]. With the possible exception of Mercury, the outer veneer of the terrestrial planets may contain other components derived from carbonaceous-chondrite-like matter and from ordinary-chondrite-like matter. Mars, for example, may have an extensive outer veneer, while for Earth, it is ≤18% by mass. Satellites may possess an internal kernel of enstatite-chondritic-matter. The particular importance of enstatite-chondritic-matter derives from the highly reduced state of oxidation during formation, which forced certain oxyphile elements, such as uranium, into the alloy portion, rather than into the silicate.

**Planetary Magnetic Fields Generated by Nuclear Fission Reactors**

Generation of magnetic fields in planets and satellites has long been conjectured to take place by the same mechanism responsible for generating Earth's magnetic field, a convecting fluid iron-alloy core dynamo. In this paper, I set forth in detail the commonality in the Solar System of the matter like that of the inside of the Earth, which is the basis for generalizing my concept of geomagnetic field generation by georeactor nuclear fission to planetocentric nuclear fission magnetic field generation in other planets and satellites. To appreciate the implications to other planets, it is beneficial to understand the circumstances related to our own planet, about which there is much more detailed information.

Elsasser [38-40] and Bullard [41] first adapted Lamor's self-exciting solar-dynamo concept [42] to explain the generation of the Earth's magnetic field. That mechanism is based upon the idea that convective motions in the Earth's fluid, electrically conducting core interacting with Coriolis forces produced by Earth's rotation, cause the fluid core to act like a dynamo, essentially a magnetic amplifier. Although being the subject of extensive investigations over more than a half century [43], there are fundamental problems with that concept.

For decades, the interior of Earth was erroneously assumed to be like an ordinary chondrite meteorite which, as was known, would have been too highly oxidized for the occurrence of radioactive elements in the core, although there was much discussion of the possibility that $^{40}$K might reside in the core. Realizing that the existence of the geomagnetic field demanded the presence of an energy source within the core, geophysicists often assumed, without corroborating evidence, that Earth's inner core was made of iron metal and that the inner core was growing, thus producing heat from the crystallization of iron supposedly to power the geomagnetic field.

I discovered that the interior of the Earth is like an enstatite chondrite [18-21] and, knowing that as a consequence of oxidation state, in enstatite chondrites uranium occurs in the alloy portion which corresponds to the Earth's core [44], I disclosed the background, feasibility and evidence of a nuclear fission georeactor at the center of the Earth as the energy source for the geomagnetic field [8-12, 45]. From fundamental considerations, as discussed above, there are only two types of primary, planet-building matter in Solar System of which only one type, like the deep interior of Earth, is capable of forming massive planetary cores. The commonality of planets with massive cores in the Solar System is indicative of their bulk compositions being of enstatite-chondrite-like matter and is the basis for my generalizing the concept of planetocentric nuclear fission reactors as planetocentric energy sources and, as discussed below, as the mechanism for generating planetary magnetic fields.

The geomagnetic field has existed for at least 3.5 billion years, as known from magnetic studies of rocks [46]. The almost universal belief that the geomagnetic field originates as a consequence of convection in the Earth's fluid core has led to little thought having been given to the possibility that there might be fundamental errors in the underlying assumptions, especially the assumption that convection in the Earth's alloy core can be sustained over extended times.

Nobel Laureate Chandrasekhar, an expert on convection [47], described convection in the following way [48]: "The simplest example of thermally induced convection arises when a horizontal layer of fluid is heated from below and an adverse temperature gradient is maintained. The adjective 'adverse' is used to qualify the prevailing temperature gradient, since, on account of thermal expansion, the fluid at the bottom becomes lighter than the fluid at the top; and this is a top-heavy arrangement which is potentially unstable. Under these circumstances the fluid will try to redistribute itself to redress this weakness in its arrangement. This is how thermal convection originates: It represents the efforts of the fluid to restore to itself some degree of stability."

As I recently noted, a fundamental difficulty arises in maintaining an adverse temperature gradient in the iron alloy fluid core [1]. Maintenance of the fundamental condition for convection stability over extended times demands maintaining an adverse temperature gradient in the core over extended times. In other words, heat continuously brought to the top of the core by convection must be continuously removed at the same rate. And, that is the problem; the silicate mantle above the core-interface is much more of an insulator than a thermal conductor. The heat capacity of the core is greater than the heat capacity of mantle silicate-rock, the thermal conductivity of the core is greater than the thermal conductivity of mantle silicate-rock, while the viscosity of the core is much less than the viscosity of mantle silicate-rock. In other words, the core is thermally well insulated by a 3,400 km-thick layer of silicate-rock.

I have described the substructure of the Earth's inner core (radius 1250 km) as having at its center the georeactor, an actinide sub-core (radius 4 km) surrounded by a fluid or slurry sub-shell (radius 6 km) composed of fission products and products of radioactive decay [10], shown for example in Figure 3 and Table 1 from [1]. The georeactor dimensions were very conservative estimates, and may in reality be as much as several times greater. The whole georeactor assembly is expected to exist at the center of Earth in contact with, and surrounded by, the nickel silicide inner core.

Convection in the fission product sub-shell is expected to be a stable feature of georeactor-like planetocentric nuclear reactors where nuclear fission produced heat is supplied directly to the base of the fission-product sub-shell whose outer boundary is a major heat sink. In Earth's georeactor, the outer boundary of the fluid sub-shell maintains contact with the semi-metallic, nickel silicide inner core, which acts as a heat sink, a thermal ballast, with reasonably good thermal conductivity to transport excess heat to the fluid iron-sulfur core, another heat sink. This arrangement enables the sub-shell's fluid to restore to itself, and to maintain an adverse temperature gradient and an enduring degree of stability. A similar arrangement would be expected for planetocentric nuclear fission reactors in general. There is some question, however, as to what observable differences might arise if the outer boundary of a fluid fission product sub-shell were in contact with a fluid planetary core in the case of a yet un-precipitated inner core.

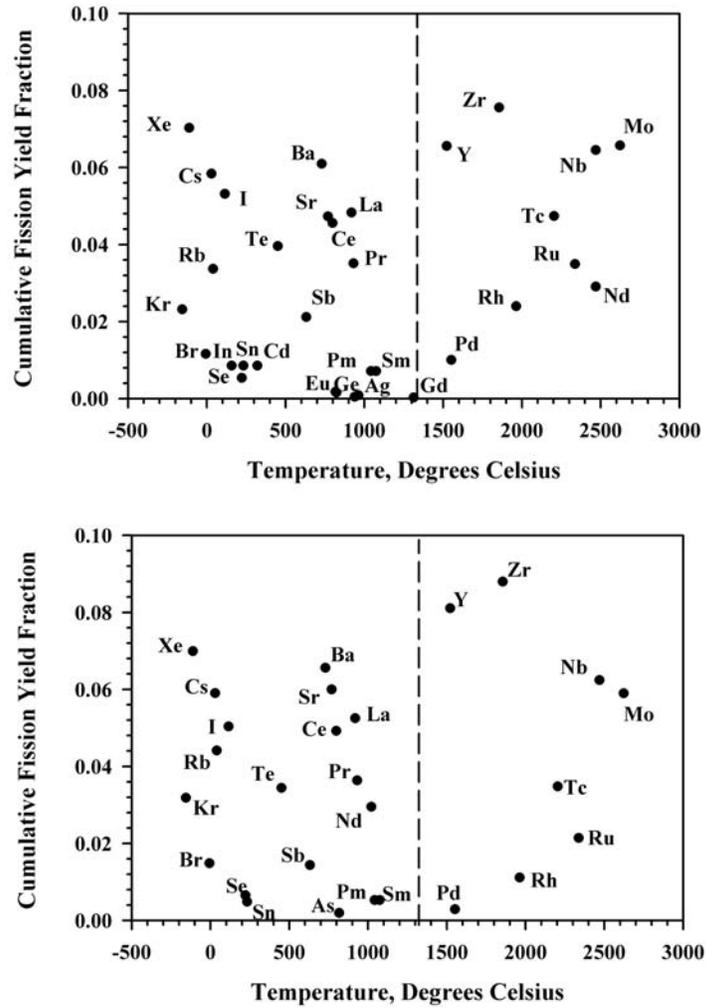

**Figure 3.** Cumulative yield fraction of $^{238}$U and $^{235}$U fast neutron fission product elements plotted vs. ambient pressure melting point of each respective element. The dashed vertical line represents the ambient pressure melting temperature of nickel silicide, $Ni_2Si$. Data from http://www.nucleonica.net. From [1].

The dynamo mechanism, thought to be responsible for generating the geomagnetic field, operates as a magnetic amplifier wherein, beginning with a small magnetic field, the combined motions of an electrically conducting fluid, driven by convection in a rotating system, amplify and maintain a more-or-less stable, much, much larger magnetic field. I have suggested for the Earth that geomagnetic field production occurs by the dynamo mechanism involving convection in the rotating, electrically conducting nuclear georeactor fission-product sub-shell, driven by nuclear fission energy produced in the georeactor sub-core [1]. This fundamentally different concept is generally applicable to magnetic field generation in planets and their satellites, and appears to obviate the seeming paradox of Mercury's magnetic field.

Mercury is composed of enstatite-chondrite-like matter, as indicated by its massive core and by reflectance spectroscopic observations showing the regolith of Mercury to be virtually devoid of FeO, like the silicates of the enstatite chondrites [49]. Paradoxically, Mercury's small size appears to preclude the existence of a fluid convecting iron-alloy core as the origin of its magnetic field. But, with Mercury's magnetic field generated by its planetocentric nuclear fission reactor, there is no paradox.

There are universal, inherent aspects to my generalized planetary dynamo concept involving convection in the rotating, electrically conducting the planetocentric nuclear fission-product sub-shell, driven by nuclear fission energy produced in the reactor sub-core. The power source and the magnetic field production mechanism are a single, self-contained unit that functions with the assurance of maintaining an adverse temperature gradient for sustained convection. By virtue of its location, operating conditions are expected to be remarkably similar, *e.g.*, the microgravity environment, despite major differences in other aspects of the planets. The presence of a seed-field is assured through the radioactive $\beta^-$ decay of neutron-rich fission products and other ionizing radiation. The generality of magnetic field generation in planets and satellites by natural nuclear fission reactors is related to the generality of enstatite-chondrite-like matter as the primary planet-building material, as shown through the fundamental considerations presented in this paper.

**Table 1**. Cumulative fission yield fractions for fast neutron fission of $^{238}$U and $^{235}$U from tabulations posted on http://www.nucleonica.net From [1].

| Element | $^{238}$U Fission Fraction | $^{235}$U Fission Fraction | Element | $^{238}$U Fission Fraction | $^{235}$U Fission Fraction |
|---|---|---|---|---|---|
| Zirconium | 0.0756 | 0.0880 | Yttrium | 0.0656 | 0.0811 |
| Xenon | 0.0703 | 0.0699 | Barium | 0.0610 | 0.0656 |
| Niobium | 0.0645 | 0.0624 | Strontium | 0.0473 | 0.0600 |
| Cesium | 0.0456 | 0.0590 | Molybdenum | 0.0658 | 0.0590 |
| Lanthanum | 0.0483 | 0.0525 | Iodine | 0.0532 | 0.0504 |
| Cerium | 0.0492 | 0.0492 | Rubidium | 0.0337 | 0.0442 |
| Praseodymium | 0.0351 | 0.0364 | Technetium | 0.0474 | 0.0348 |
| Tellurium | 0.0396 | 0.0344 | Krypton | 0.0231 | 0.0318 |
| Neodymium | 0.0291 | 0.0295 | Ruthenium | 0.0349 | 0.0215 |
| Bromine | 0.0116 | 0.0148 | Antimony | 0.0212 | 0.0144 |
| Rhodium | 0.0240 | 0.0111 | Selenium | 0.00545 | 0.0065 |
| Samarium | 0.0071 | 0.0053 | Promethium | 0.0071 | 0.0053 |
| Tin | 0.0085 | 0.0048 | Palladium | 0.0100 | 0.0029 |
| Arsenic | 0.0018 | 0.0020 | Europium | 0.0015 | 0.0008 |
| Indium | 0.0085 | 0.0007 | Silver | 0.0008 | 0.0006 |
| Cadmium | 0.0085 | 0.0005 | Germanium | 0.0005 | 0.0005 |
| Gadolinium | 0.0003 | 0.0001 | Gallium | 0.0001 | 0.0001 |